\documentstyle[11pt,newpasp,twoside,epsf]{article}
\def\edcomment#1{\iffalse\marginpar{\raggedright\sl#1\/}\else\relax\fi}
\reversemarginpar

\begin{document}
\title{Lensing results from the Red-sequence Cluster Survey}
\author{Henk Hoekstra$^{1,2}$, Howard Yee$^{2}$, \& Mike Gladders$^{2,3}$}
\affil{$^1$ CITA, University of Toronto, Toronto, Ontario M5S 3H8, Canada\\
$^2$ Department of Astronomy and Astrophysics, University of Toronto,
Toronto, Ontario M5S 3H8, Canada\\
$^3$ Observatories of the Carnegie Institution of Washington,
813 Santa Barbara Street, Pasadena, California 91101}

\begin{abstract}

We present a variety of weak lensing results based on the ongoing analysis
of $R_C$-band imaging data from the Red-Sequence Cluster Survey (RCS). 
We briefly discuss the weak lensing signal induced by intervening
large scale structure (cosmic shear), and study the properties
of the dark matter halos surrounding galaxies with $19.5<R_C<21$
(which have a median redshift of $z=0.35$) using a parametrized
mass model for the galaxy mass distribution. This allows us for the
first time to constrain the extent of the halos. We find a robust
upper limit for the truncation parameter $s<470 h^{-1}$ kpc (99.7\%
confidence). We also study the biasing properties of these galaxies as
a function of scale. We find that both the bias parameter $b$ and the
galaxy-mass cross-correlation coefficient $r$ vary with scale (on
scales $0.1-6~h^{-1}$ Mpc). Interestingly, we find that 
$r=0.57^{+0.08}_{-0.07}$ on scales $\sim 0.5 h^{-1}$ Mpc.

\end{abstract}

\section{Introduction}

The Red-Sequence Cluster Survey\footnote{\tt
http://www.astro.utoronto.ca/${\tilde{\ }\!}$gladders/RCS} (e.g.,
Gladders \& Yee 2000) is the largest area, moderately deep imaging
survey ever undertaken on 4m class telescopes. The planned survey
comprises 100 square degrees of imaging in 2 filters (22 widely
separated patches imaged in $R_C$ and $z'$), and will
provide a large sample of optically selected clusters 
of galaxies with redshifts $0.1<z<1.4$.

The survey allows a variety of studies, such as constraining cosmological
parameters from the measurement of the evolution of the number density
of galaxy clusters as a function of mass and redshift, and studies
of the evolution of cluster galaxies, blue fraction, etc. at redshifts
for which very limited data are available at present.

The data are also useful for a range of lensing studies. Strong
lensing by clusters of galaxies allows a detailed study of their core
mass distribution. In addition, given the shallowness of the survey,
the arcs are sufficiently bright to be followed up spectroscopically
(Gladders, Yee, \& Ellingson 2001). Thanks to the large magnifications
of the arcs, the survey allows unprecedented studies of the properties
of high redshift galaxies. Furthermore, in combination with detailed
modeling of the cluster mass distribution, the geometry of the images
can be used to constrain $\Omega_m$.

In these proceedings we concentrate on some of the weak lensing
applications. A detailed description of the data and the weak lensing
analysis is presented in Hoekstra et al. (2002a).  A careful
examination of the residuals suggests that the object analysis, and
the necessary corrections for observational distortions work well,
which allows us to obtain accurate measurements of the weak lensing
signal.

\section{Measurement of Cosmic Shear}

The weak distortions of the images of distant galaxies by intervening
matter provide an important tool to study the projected mass
distribution in the universe and constrain cosmological parameters
(e.g., van Waerbeke et al. 2001). Compared to other studies, the RCS
data are relatively shallow, resulting in a lower cosmic shear
signal. However, the redshift distribution of the source galaxies,
which is needed to interpret the results, is known fairly well.  

We use the photometric redshift distribution inferred from the Hubble
Deep Field North and South to compare the observed lensing signal to
CDM predictions.  For an $\Omega_m=0.3$ flat model we obtain
$\sigma_8=0.81^{+0.14}_{-0.19}$ (95\% confidence), in good
agreement with the measurements of van Waerbeke et al. (2001).
A detailed discussion of this measurement is presented in
Hoekstra et al. (2002a).

\begin{figure}[h!]
\centering 
\leavevmode 
\hbox{
\epsfxsize=7.5cm
\epsfbox{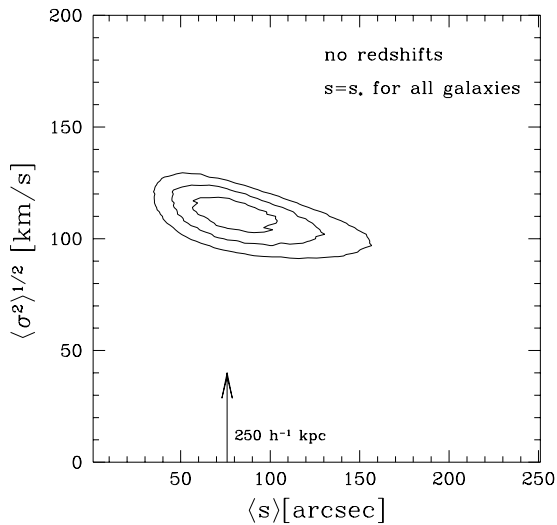}}
\caption{\small Likelihood contours for the mass
weighted velocity dispersion, and the average value of the truncation
parameter $s$ for galaxies with $19.5<R_C<21$. We have also indicated 
the physical scale when $s$ is the same for all galaxies. The contours 
indicate 68.3\%, 95.4\%, and 99.7\% confidence limits on two parameters 
jointly.}
\end{figure}

\section{Galaxy-Galaxy Lensing}

Weak lensing is also an important tool to study the dark matter halos
of field (spiral) galaxies. The weak lensing signal can be measured
out to large projected distances. Hence it provides a powerful probe
of the gravitational potential at large radii.  Unfortunately, the
lensing signal induced by an individual galaxy is too low to be
detected, and one can only study the ensemble averaged signal around a
large number of lenses.

The results presented here are based on 16.4 deg$^2$ of CFHT data.
We use galaxies with $19.5<R_C<21$ as lenses, and galaxies with
$21.5<R_C<24$ as sources which are used to measure the lensing
signal. The redshift distribution of the lenses is
known spectroscopically from the CNOC2 field galaxy redshift survey
(e.g., Yee et al. 2000), and for the source redshift distribution we
use the photmetric redshift distribution from the HDF North and
South. The adopted redshift distributions give a median redshift
$z=0.35$ for the lens galaxies, and $z=0.53$ for the source galaxies.

We use a parametrized mass model with a (smoothly) ``truncated'' halo
to study the properties of the dark matter halos surrounding the
sample of lens galaxies. The results are presented in Figure~1. With
the adopted redshift distribution we obtain
$\langle\sigma^2\rangle^{1/2}=111\pm5$ km/s. It turns out that the
quoted value is close to that of an $L^*$ galaxy, and our results are
in fair agreement with other estimates.

In addition, for the first time, the average extent of the dark matter
halo has been measured. Under the assumption that all halos have the
same truncation parameter, we find a 99.7\% confidence upper limit of
$\langle s\rangle <470 h^{-1}$ kpc.  More realistic scaling relations
for $s$ give lower values for the physical scale of $\langle
s\rangle$, and therefore the result presented here can be interpreted
as a robust upper limit.

\begin{figure}[h!]
\leavevmode
\centering
\hbox{
\epsfysize=7.5cm
\epsfbox{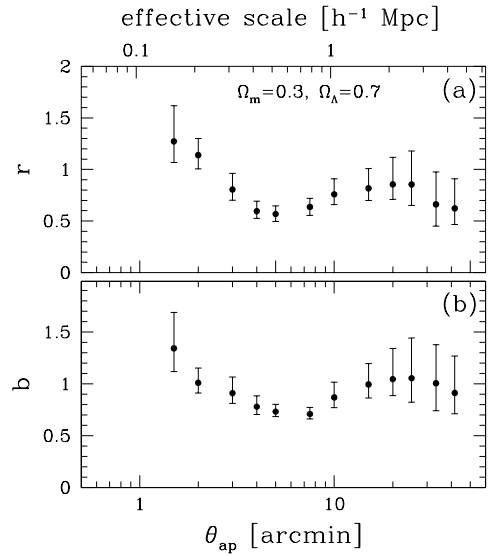}}
\vspace{-0.3cm}
\caption{\small (a) The measured value of the galaxy-mass
cross correlation coefficient $r$ as a function of scale for the
$\Lambda$CDM cosmology. (b) The bias parameter $b$ as a function
of scale. The upper axis indicates the effective physical scale probed by the
compensated filter at the median redshfit of the lenses
$(z=0.35)$. The errorbars correspond to the 68\% confidence intervals.
Note that the measurements at different scales are slightly
correlated.
\label{bias}}
\end{figure}

\section{Measurement of Galaxy Biasing}

The relation between the galaxy distribution and the
dark matter distribution (i.e., galaxy biasing) is intimately
linked to the details of galaxy formation. Most current constraints
come from dynamical measurements that probe relatively large scales
($\ge$ a few Mpc). Weak lensing provides a unique and powerful
way to study the biasing properties of galaxies on smaller scales.

Through weak lensing we can measure the bias parameter $b$ (which
corresponds to the linear regression of the mass density contrast and
the galaxy density contrast) and the galaxy-mass cross-correlation
coefficient $r$ (which mixes non-linear and stochastic biasing) as a
function of scale.

The RCS alone provides an accurate measurement of the ratio $b/r$,
and the first results were published in Hoekstra et al. (2001).
However, combining the RCS measurements with the results from
the VIRMOS-DESCART survey (van Waerbeke et al. 2001), allows
a determination of $b$ and $r$ separately. The results
presented in Figure~2 are based on 45 deg$^2$ of RCS data and
6.5 deg$^2$ of VIRMOS-DESCART data.

We find that $b$ changes with scale for our sample of lens galaxies
($19.5<R_C<21$), which have luminosities around $L^*$. For the
currently favored cosmology $(\Omega_m=0.3,~\Omega_\Lambda=0.7)$, we
find $b=0.71^{+0.06}_{-0.05}$ (68\% confidence) on a scale of $0.5-1
h^{-1}$ Mpc, increasing to $\sim 1$ on larger scales.  The value of
$r$ hardly depends on the assumed cosmology, and we find that $r\sim
1$ on scales less than $250h^{-1}$ kpc. Hence, we ``find'' a halo
around every (massive) galaxy. However, on larger scales ($\sim 0.5
h^{-1}$ Mpc) $r$ is significantly lower than unity (we find a minimum
value of $r=0.57^{0.08}_{-0.07}$), thus suggesting significant
stochastic biasing and/or non-linear biasing.  On even larger scales
($>2 h^{-1}$Mpc) the value of $r$ is lower, but consistent with $r=1$.
As $r$ is linked intimately to the details of galaxy formation, our
results provide unique constraints on such models. A detailed
discussion of these measurements can be found in Hoekstra et
al. (2002b).


\begin{references}
\reference
	Gladders, M.D., Yee, H.K.C. 2000, to appear in ``The New Era
	of Wide-Field Astronomy'' (astro-ph/0011073)
\reference
	Gladders, M.D., Yee, H.K.C., Ellingson, E. 2001, AJ, in press
\reference
	Hoekstra, H., Yee, H.K.C., \& Gladders, M.D. 2001, ApJ, 558, L11
\reference
	Hoekstra, H., Yee, H.K.C., Gladders, M.D., Barrientos, L.F.
	Hall, P.B., \& Infante, L. 2002a, ApJ, submitted
\reference
	Hoekstra, H., van Waerbeke, L., Gladders, M.D., Mellier, Y., 
	\& Yee, H.K.C. 2002b, ApJ, submitted
\reference
	van Waerbeke, L., et al. 2001, A\& A, 374, 757 (astro-ph/0101511)
\reference
	Yee, H.K.C. et al. 2000, ApJS, 129, 475

\end{references}
\end{document}